\documentclass{aa}  
\usepackage{color}
\usepackage{graphicx}
\usepackage{lscape}
\usepackage{natbib}
\usepackage{natbib,twoopt}
\usepackage[varg]{txfonts}
\usepackage{url}
\usepackage{xspace}
\usepackage{amssymb}
\usepackage{stmaryrd}
\usepackage{siunitx}
\usepackage{textcomp}
\usepackage{pifont}

\usepackage[colorlinks = true,
linkcolor= blue,
citecolor = blue,
filecolor = blue,
urlcolor = blue,
]{hyperref}

\usepackage[usenames,dvipsnames]{xcolor}

\bibpunct{(}{)}{;}{a}{}{,}

\newcommand{\jn}{\object{J0546$+$0836}\xspace}
\newcommand{\jnz}{\object{J0927$-$6335}\xspace}
\newcommand{\jnd}{\object{J1332$-$3541}\xspace}

\newcounter{Rco}

\newcommand{\logg}{\mbox{$\log g$}\xspace}
\newcommand{\loggw}[1]{\mbox{$\log g\hspace{-0.5mm} =\hspace{-0.5mm}  #1$}}

\newcommand{\Teff}{\mbox{$T_\mathrm{eff}$}\xspace}

\newcommand{\Lsol}{$L_\odot$}
\newcommand{\Msol}{$M_\odot$}
\newcommand{\Rsol}{$R_\odot$}
\newcommand{\Mdot}{$\dot{M}$}

\defcitealias{El-Badry2023}{E23}

\begin{document}

\title{The photospheres of the hottest fastest stars in the Galaxy}

\author{Klaus Werner\inst{1} \and Nicole Reindl\inst{2} \and Thomas Rauch\inst{1} \and Kareem El-Badry\inst{3} \and Antoine B\'edard\inst{4}}

\institute{Institut f\"ur Astronomie und Astrophysik, Kepler Center for
  Astro and Particle Physics, Eberhard Karls Universit\"at, Sand~1, 72076
  T\"ubingen, Germany\\ \email{werner@astro.uni-tuebingen.de} 
\and
  Landessternwarte Heidelberg, Zentrum f\"ur Astronomie, Ruprecht-Karls-Universit\"at, Königstuhl~12, 69117 Heidelberg, Germany
\and
Department of Astronomy, California Institute of Technology, 1200 East California Boulevard, Pasadena, CA 91125, USA
\and
Department of Physics, University of Warwick, Coventry CV4 7AL, UK
}

\date{Received 16 October 2023 / Accepted 16 November 2023}

\authorrunning{K. Werner et al.}
\titlerunning{Hottest fastest stars in the Galaxy}

\abstract{
We perform nonlocal thermodynamic equilibrium (NLTE) model atmosphere analyses of the three hottest hypervelocity stars (space velocities between $\approx$ 1500--2800 km\,s$^{-1}$) known to date, which were recently discovered spectroscopically and identified as runaways from Type Ia supernovae. The hottest of the three (\jn, effective temperature \Teff = 95\,000 $\pm$ 15\,000\,K, surface gravity \logg = $5.5 \pm 0.5$) has an oxygen-dominated atmosphere with a significant amount of carbon (C = $0.10 \pm 0.05$, O = $0.90 \pm 0.05$, mass fractions). Its mixed absorption+emission line spectrum exhibits photospheric absorption lines from \ion{O}{v} and \ion{O}{vi} as well as \ion{O}{iii} and \ion{O}{iv} emission lines that are formed in a radiation-driven wind with a mass-loss rate of the order of \Mdot = $10^{-8}$ \Msol yr$^{-1}$. Spectroscopically, \jn is a [WC]--PG1159 transition-type pre-white dwarf. The second object (\jnz) is a PG1159-type white dwarf with a pure absorption-line spectrum dominated by \ion{C}{iii}/\ion{C}{iv} and \ion{O}{iii}/\ion{O}{iv}. We find \Teff = 60\,000 $\pm$ 5000\,K, \logg = $7.0 \pm 0.5$, and a carbon- and oxygen-dominated atmosphere with C = $0.47 \pm 0.25$, O = $0.48 \pm 0.25$, and possibly a minute amount of helium (He = $0.05 \pm 0.05$). Comparison with post-AGB evolutionary tracks suggests a mass of $M\approx0.5$\,\Msol\, for both objects, if such tracks can safely be applied to these stars. We find the third object (\jnd ) to be a relatively massive ($M=0.89$\,\Msol) hydrogen-rich (DAO) white dwarf with \Teff = 65\,657 $\pm$ 2390\,K, \logg = $8.38 \pm 0.08$, and abundances H = $0.65 \pm 0.04$ and He = $0.35 \pm 0.04$. We discuss our results in the context of the ``dynamically driven double-degenerate double-detonation'' (D$^6$) scenario proposed for the origin of these stars.
}


\maketitle
%

\section{Introduction}
\label{sect:intro}

The gravitational-wave-driven inspiral of two white dwarfs (WDs) in a close binary has long attracted interest as a progenitor channel for Type Ia supernovae \citep[e.g.,][]{Webbink1984, Livne1990, Guillochon2010}. In several subclasses of such ``double-degenerate'' models, only one of the two WDs explodes \citep[e.g.,][]{Papish2015, Pakmor2022, Burmester2023}. The other WD, suddenly free from the 
gravity of its companion, escapes at roughly its orbital velocity, which can range from 1000 to 2500\,$\rm km\,s^{-1}$ \citep[e.g.,][]{Bauer2021}. This surviving WD is predicted to have unusual surface abundances, because the hydrogen and/or helium in its outer layers is stripped off by the companion both prior to and during the explosion, and also because ejecta from the companion may be deposited on its surface during the supernova and remain bound. Such a surviving WD could be identified as a ``living remnant'' of the supernova long after the bubble formed during the explosion has dissipated back into the interstellar medium. 

Several hypervelocity WDs with unusual abundances have been discovered in recent years. Using data from {\it Gaia} Data Release 2 (DR2), \citet{Shen2018} identified three objects with space velocities of higher than $\approx 1000\,\rm km\,s^{-1}$, radii of $0.1-0.3\,R_{\odot}$, and effective temperatures of $T_{\rm eff} \approx 7000\,{\rm K}$, placing them between WDs and main sequence stars in the HR diagram. No detailed abundance analysis has been performed on these objects, but their atmospheres appear to be free of hydrogen and rich in metals \citep[e.g.,][]{Chandra2022}. More recently, \citet[][hereafter E23]{El-Badry2023} reported the discovery of four hypervelocity WDs that appear to be hotter and smaller than the objects discovered by \citet{Shen2018}, but have similarly high ---and, in some cases, even higher--- space velocities. Both these hot objects and their cooler, puffier cousins have been referred to as ``D$^6$'' stars, after the ``dynamically driven, double-degenerate, double-detonation'' model for their formation \citep{Shen2018b}. While other classes of models for thermonuclear explosions producing high-velocity WDs have been explored \citep{Raddi2019, Jones2019, Gansicke2020}, the D$^6$ scenario is currently the only model predicting velocities above $1000\,\rm km\,s^{-1}$.

A detailed spectral analysis of candidate D$^6$ stars has not been carried out anywhere in the literature. \citetalias{El-Badry2023} compared the optical spectra of their candidates to LTE model spectra calculated with the general-purpose opacity sampling and spectral synthesis codes ATLAS12 and SYNTHE \citep{Kurucz1970SAOSR, kurucz_model_1979, Kurucz1992}, but the authors cautioned that these codes were not designed for hot, high-gravity stars, their grid of models was not exhaustive, and the strengths of many lines included in their models were uncertain. While the model spectra they produced matched the observed spectra well enough for the measurement of radial velocities, many lines were left unidentified. It is therefore highly desirable to analyze the spectra of these unusual objects in detail with a code tailored to hot WDs. That is the goal of this work.

The remainder of this paper is organized as follows. In Sect.\,\ref{sect:models}, we introduce our model atmospheres and the analysis strategy. In Sections \ref{sect:jn}--\ref{sect:jnd}, for each of our three program stars, we describe the spectral line identification and the spectral fitting with our models in detail, and present the resulting determinations of effective temperature (\Teff), surface gravity ($g$), and chemical composition. In Sect.\,\ref{sect:sed}, we derive the stellar parameters mass ($M$), radius ($R$), and luminosity ($L$) using theoretical stellar evolutionary tracks, and determine their spectroscopic distances by fitting spectral energy distributions (SEDs). We summarize and discuss our results in Sect.\,\ref{sect:summary}.

\section{Model atmospheres}
\label{sect:models}

We used the T\"ubingen Model-Atmosphere Package (TMAP) to compute nonlocal thermodynamic equilibrium (NLTE), plane-parallel, line-blanketed atmosphere models in radiative and hydrostatic equilibrium \citep{2003ASPC..288...31W}.
For the two carbon- and oxygen-dominated stars (\jn and \jnz), we computed models of the type introduced in detail by \citet{2014A&A...569A..99W}, which were tailored to investigating the optical spectra of relatively cool PG1159 stars (\Teff $\approx$ 90\,000\,K, \logg = 7.5); that is, hydrogen-deficient WDs resulting from a late He-shell flash \citep{2006PASP..118..183W}. Such models were also successfully used to analyze recently discovered hot subdwarfs (\Teff $\approx$ 55\,000\,K, \logg $\approx 5.5$) that are covered by helium-burning ash resulting from a binary--white dwarf merger \citep{2022MNRAS.511L..66W}. The chemical constituents of the models are helium, carbon, and oxygen. We found the model calculations to be rather cumbersome because of numerical instabilities. It was only possible to achieve convergence with careful treatment and it was therefore impossible to compute model grids to determine the atmospheric parameters. We computed a series of models in order to find a good fit to the observed line spectra. 

For the third examined object (\jnd), we used grids of WD model atmospheres composed of hydrogen and helium, computed with the TMAP code by \cite{Reindl+2023}. In order to determine \Teff, \logg, and the H/He abundance ratio, we performed a global $\chi^2$ spectral fit considering several absorption lines of hydrogen and helium. A fourth object discovered by \citetalias{El-Badry2023}, J1235-3752, is not analyzed here because its lower temperature ($T_{\rm eff}\approx 21\,000\,{\rm K}$ according to \citetalias{El-Badry2023}) makes it unsuitable for analysis with TMAP.

\begin{figure*}
 \centering  \includegraphics[width=1.00\textwidth]{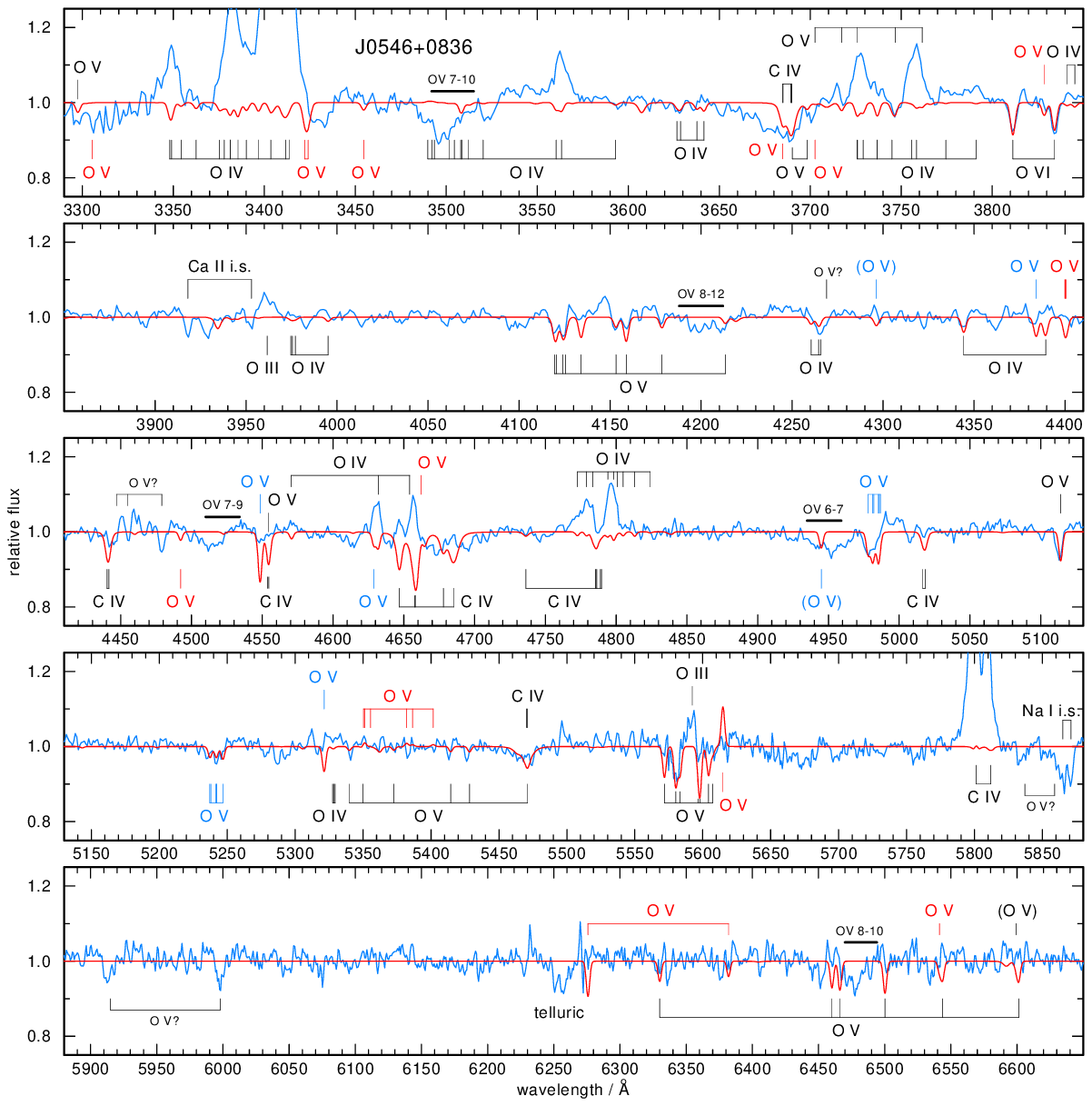}
 \caption{Keck/LRIS spectrum of \jn shifted to rest wavelength and compared to our final model (red graph) with \Teff
   = 95\,000\,K, \logg = 5.5, O = 0.90, and C = 0.10 (mass
   fractions). The spectrum is dominated by oxygen lines from \ion{O}{iv} in emission (formed in a stellar wind and therefore not reproduced by our static model) and \ion{O}{v} in absorption, plus the \ion{O}{vi} absorption doublet at 3411/34\,\AA\ and two wind-emission lines of \ion{O}{iii} (3962 and 5592\,\AA). We note, in particular, the (highly excited) Rydberg lines from \ion{O}{v} (which are not modeled) indicated by the thick horizontal black bars with respective principal quantum numbers. Small \ion{O}{v} labels with question marks indicate probable \ion{O}{v} lines in the observation that are nevertheless not included in the model atom. See the main text for the meaning of the colors of the \ion{O}{v} line labels. The only other species identified is carbon; e.g., the \ion{C}{iv} multiplet at 3690\,\AA. The \ion{C}{iv} emissions (e.g., the doublet at 5801/12\,\AA) are also formed in the wind. }
\label{fig:J0546+0836}
\end{figure*}

\section{J0546+0836}
\label{sect:jn}

We analyzed the rectified spectrum of \jn presented by \citetalias{El-Badry2023} in the top panel of their fig.\,3. The spectrum was obtained on MJD 59959.35 with the Low Resolution Imaging Spectrometer (LRIS) on the 10\,m Keck-I telescope on Maunakea, Hawaii. The spectral resolution is $R \approx 1500$. Our model spectra were folded accordingly with a Gaussian. Figure\,\ref{fig:J0546+0836} shows the LRIS spectrum, shifted to rest wavelength assuming a radial velocity of 1350 km\,s$^{-1}$. \citetalias{El-Badry2023} identified only a few of the spectral lines and concluded that the star is free of helium and dominated by carbon and oxygen. We have identified almost all of the features and find that the line spectrum is dominated by oxygen. Some carbon features are also present. 

First, we note that the spectrum contains numerous emission lines. \citetalias{El-Badry2023} noted that several of these lines resemble those of PG 1159 stars, where they originate in the photosphere due to NLTE effects. However, as discussed below, our analysis suggests that the emission lines originate in a stellar wind, as in [WC] stars.

The oxygen line spectrum is a mixture of low-ionization emission lines (\ion{O}{iii} and \ion{O}{iv}) that are formed in the wind and high-ionization absorption lines (\ion{O}{v} and \ion{O}{vi}) that are formed in the photosphere. We identified wind emission lines mainly from \ion{O}{iv}, in particular the strongest emission at 3410\,\AA. We also see emission lines from \ion{O}{iii}, namely at 3962 and 5592\,\AA, which are relatively weak compared to the bulk of the \ion{O}{iv} lines. Photospheric absorption lines mainly stem from \ion{O}{v} (e.g., at 5114\,\AA). \citetalias{El-Badry2023} already identified the \ion{O}{vi} $3s-3p$ doublet at 3811/34 in absorption. These are the only \ion{O}{vi} lines present because, as we show below, the temperature of the star is too low to exhibit other (highly excited) lines from this ion. \citetalias{El-Badry2023} attributed the broad absorption feature at 3500\,\AA\ to \ion{O}{vi}, but our findings suggest it is one of several Rydberg lines from \ion{O}{v} (see below). 

The oxygen line spectrum of \jn allows a coarse assessment of the effective temperature, and the result served as a starting point for our spectral fitting. We compared the mixed absorption+emission line spectrum of \jn to other hydrogen-deficient hot stars that are either dominated by emission lines or absorption lines. The relative strengths of the emission lines from \ion{O}{iii} and \ion{O}{iv} are reminiscent of Wolf-Rayet central stars of planetary nebulae of intermediate spectral type. These low-mass stars belong to the [WC] sequence. Most [WC] stars have been classified as [WC]-early or [WC]-late, with only a few intermediate types \citep[e.g.,][]{1993A&AS..102..595T}. One of them for instance is LMC-SMP~61, which was assigned the spectral type [WC 5--6] and was analyzed in detail by \cite{2004A&A...413..329S}. Like \jn, it shows predominantly emission lines from \ion{O}{iii} and \ion{O}{iv}. Emissions from the higher ionization stages \ion{O}{v} and \ion{O}{vi} are rather weak. In \jn, such higher ionization lines are also present (albeit in absorption), meaning that the oxygen ionization balance in both stars is similar. The stellar temperature measured for LMC-SMP~61 is $T_\star$ = 87\,500\,K.\footnote{The quantity $T_\star$ for stars with a massive wind is defined at a particular Rosseland optical depth (here: $\tau_{\rm R}=20$) and is related, but not identical, to the effective temperature in stellar photospheres.} Hot hydrogen-deficient stars exhibiting predominantly absorption line spectra are the PG1159 stars, which are very hot (pre-) WDs and are thought to be the progeny of [WC] stars. Similar relative strengths of the absorptions of the \ion{O}{vi} doublet at 3811/34\,\AA\ and the \ion{O}{v} singlet at 5114\,\AA\ in \jn are observed in PG1159 stars in the temperature range $\approx$100\,000--120\,000\,K \citep[PG\,2131$+$066 and BMP\,J0739$-$1418;][]{2014A&A...569A..99W,2023A&A...676A...1W}.   The \ion{O}{v} line and the \ion{O}{vi} doublet disappear at higher and lower temperatures, respectively. Therefore, we can conclude that the effective temperature of \jn is near 100\,000\,K. 

The facts that \jn exhibits a mixed absorption+emission line spectrum and that the strengths of its emission lines are much weaker than in [WC] stars must be the consequence of a weak wind with a mass-loss rate considerably lower than that of [WC] stars (that of LMC-SMP~61 is \Mdot = $10^{-6.12}$ \Msol yr$^{-1}$). There are a two objects regarded as [WC]--PG1159 transition types exhibiting similar mixed absorption+emission spectra to \jn, but these have higher effective temperatures \citep[the central stars of the planetary nebulae Abell~30 and Abell~78, \Teff = 115\,000\,K;][]{2015ApJ...799...67T,2015ASPC..493..141T}. The presence of \ion{O}{vi} emission lines and lack of \ion{O}{iv} emission lines in their spectra, in contrast to \jn, again points at a lower temperature of the latter, of around 100\,000\,K. Their mass-loss rates are $10^{-7.8}$ \Msol yr$^{-1}$ and that value should be representative of \jn as well. Ultraviolet spectroscopy will be necessary to characterize the wind of \jn. We expect prominent P~Cygni line profiles from \ion{C}{iv}~1550\,\AA\ and \ion{O}{v}~1371\,\AA\ for example.

The only other chemical element that can be identified in the spectrum is carbon, namely by \ion{C}{iv} lines. \citetalias{El-Badry2023} already identified two sets of emission lines from \ion{C}{iv}: the strong $3s-3p$ doublet at 5801/12\,\AA, and a weaker line at 7724\,\AA. The strong and broad emission profile of 5801/12\,\AA\ is also seen in [WC] stars and the line is formed in the stellar wind. Other \ion{C}{iv} lines that are usually observed in absorption in PG1159 stars of similar temperature are also affected by the wind because they appear either in emission (at 4647/4658\,\AA) or are filled in by emission such that they are not detectable (e.g., at 4441\,\AA). The only clear photospheric \ion{C}{iv}  absorption feature is at 3690\,\AA. However, this latter also could be affected by the wind. The absorption blueward of the line core could indicate a P~Cygni-like profile for this line. The blueward extension of about 25\,\AA\ would indicate a wind velocity of about 2000 km\,s$^{-1}$, which is similar to the value measured for the [WC] star LMC-SMP~61 \cite[1400 km\,s$^{-1}$,][]{2004A&A...413..329S}.

We also observe a strong absorption feature from the interstellar NaD doublet that, because of the radial velocity correction of the spectrum, appears blueshifted by about 26\,\AA\ to 5870\,\AA. Interstellar \ion{Ca}{ii} H and K lines are also observed and labeled.
In the following subsections, we describe  the oxygen line identifications in detail, ion by ion, and we refer to them when performing our model atmosphere fitting.

\subsection{\ion{O}{iii} and \ion{O}{iv} lines}

Two emission lines located at 3962 and 5592\,\AA\ are isolated \ion{O}{iii} transitions. They are singlets between sublevels with principal quantum number $n=3$ and are both marked with large line intensities in the NIST\footnote{\url{https://www.nist.gov/pml/atomic-spectra-database}} atomic database.
The majority of the observed emission lines stem from \ion{O}{iv}. They are transitions within the doublet and quartet systems. Quartet lines involve $n=3$ sublevels with a single excited electron that has excitation energies in the narrow range of $E$ = 439\,000--504\,000 cm$^{-1}$. In the spectrum of J0546+0836, these quartet lines are responsible for the strongest oxygen emission features located at 3381--3426, 3726--3774, and 4772--4801\,\AA. From the doublet system, we see two relevant transitions between $n=3$ sublevels with a single excited electron, namely one at 3404--3414\,\AA,\ which is contributing to the strongest oxygen emission line and the other at 3348--3349\,\AA,\ which is causing an observed emission line. All other detected doublet lines involve doubly or singly excited levels with $n>3$. Respective level energies are higher and are within the range of $E$ = 511\,000--597\,000 cm$^{-1}$. Transitions with $n=5\rightarrow6$ might be responsible for weak absorption lines, namely at 3627/29\,\AA\ ($5p-6d$) and 4344\,\AA\ ($5f-6g$), but may also be driving weak emissions, namely at 4389\,\AA\ ($5d-6f$) and 4570\,\AA\ ($5f-6d$).

\begin{figure*}
 \centering  \includegraphics[width=1.00\textwidth]{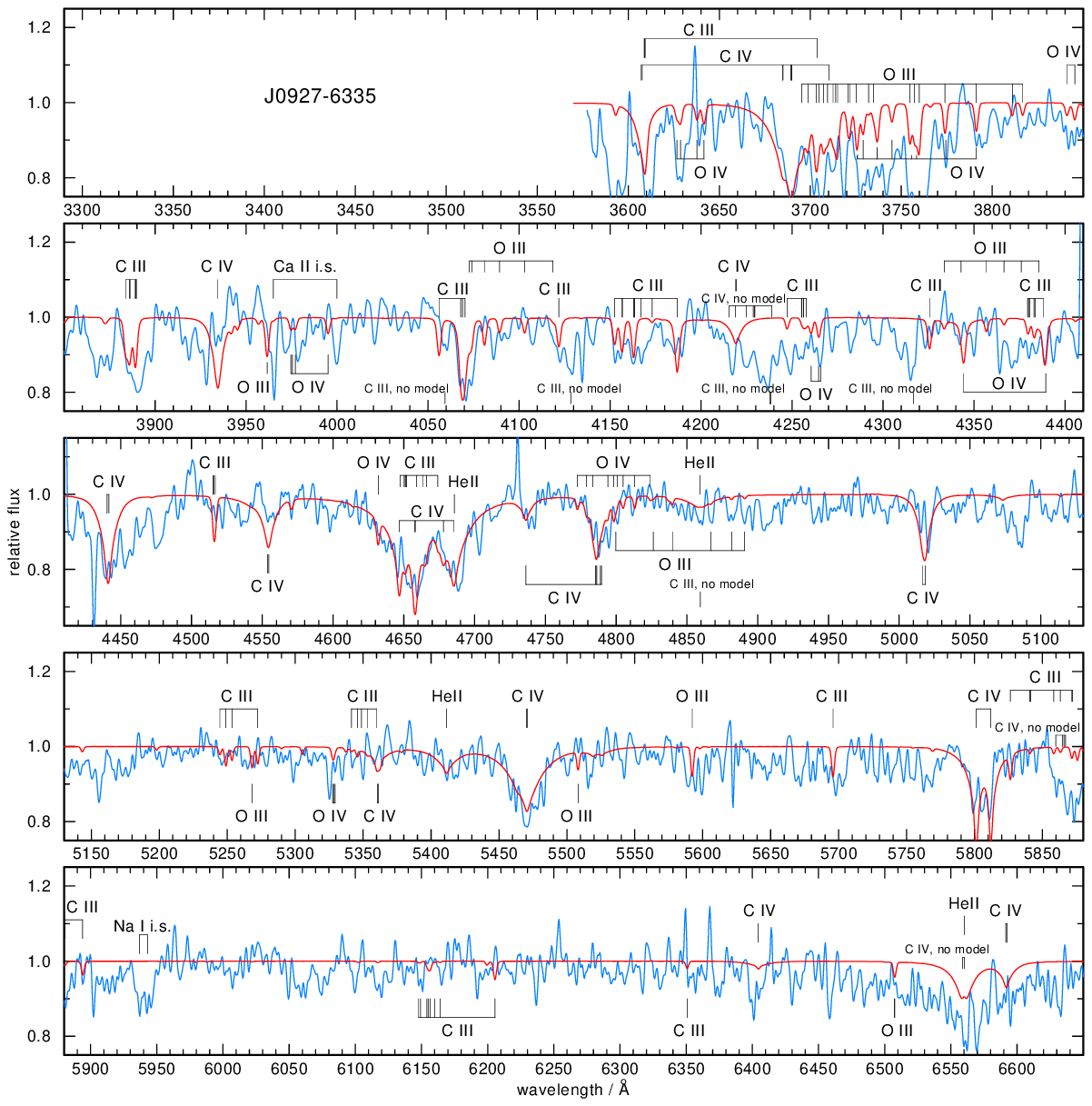}
 \caption{MagE spectrum of \jnz shifted to rest wavelength and compared to our final model (red graph) with \Teff
   = 60\,000\,K, \logg = 7.0, C = 0.47, O = 0.48, and He = 0.05 (mass
   fractions). The observed spectrum is dominated by \ion{C}{iii}/\ion{C}{iv} and \ion{O}{iii}/\ion{O}{iv} lines. The absorption trough at 4620--4710\,\AA\ is similar to that seen in PG1159 stars, but there is no contribution of \ion{He}{ii} in the case of \jnz. Line labels including ``no model'' means that the respective lines are not computed in the synthetic spectrum. In particular, the broad absorption feature at 6560\,\AA\ is caused by \ion{C}{iv} $n=8-12$ transitions that are not included in our NLTE model atom.}
\label{fig:J0927-6335}
\end{figure*}

\subsection{\ion{O}{v} and \ion{O}{vi} lines}

The ground state of \ion{O}{v} has the electron configuration 1s$^2$2s$^2$; thus, we have a singlet and a triplet system. Excited levels involve either one or two excited electrons, namely 2s\,nl or 2p\,nl.
The lines from \ion{O}{v} in the observed wavelength region fall into three main groups. 
\begin{enumerate}[i.]
    \item  The first group comprises transitions between sublevels with $n=3$, which are designated by black labels in Fig.\,\ref{fig:J0546+0836}. In comparison to the levels of the second group of lines, they have relatively low excitation energies within $E$ = 546\,000--720\,000 cm$^{-1}$.
    \item  The second group consists of lines involving highly excited levels with $n>5$ and $\Delta n>0$ 
($E>$ 838\,000 cm$^{-1}$), indicated by blue labels in Fig.\,\ref{fig:J0546+0836}. Some of them are probably affected by linear Stark broadening because levels with different angular quantum number with the same $n$ have similar excitation energies, which are close to the hydrogen-like case. We even identified Rydberg lines between very highly excited states ($n$ = 6, 7, 8 $\rightarrow$ $n'$ = 7, 9, 10, 11, 12). Their positions are indicated by the horizontal black bars in Fig.\,\ref{fig:J0546+0836}. These are clearly broad features affected by linear Stark broadening. To our knowledge, this is the first identification of \ion{O}{v} Rydberg lines in any astrophysical spectrum. The lines of these two groups are transitions between levels with either singly or doubly excited valence electrons.
\item  The third group of lines, indicated by red labels in Fig.\,\ref{fig:J0546+0836}, are transitions from a singly excited to a doubly excited level or vice versa, involving excitation energies $E>$ 808\,000 cm$^{-1}$.
\end{enumerate}

Our analysis predominantly relies on the lines of group (i). Because we do not know how to treat the (possibly important) linear Stark broadening of profiles of lines from group (ii), care must be taken when we compare them to observations. Lines from group (iii) are found to be much too strong in the models compared to the observation (e.g., at 6276\,\AA) or are even in emission (at 5615\,\AA), which is also not observed; the reason for this is unknown. One possibility is that the calculation of electron collisional rates between singly and doubly excited levels is poorly approximated by the formula from \cite{1962ApJ...136..906V}, which we use in our model atom.
A few other absorption features in the observed spectrum probably stem from lines that are not included in our model atom. These are indicated by labels with a question mark (``\ion{O}{v}?'').


The $3s-3p$ doublet of \ion{O}{vi} located at 3811/34\,\AA\ is seen in absorption. The excitation energy of the lower level is about $E$ = 640\,000 cm$^{-1}$. All other \ion{O}{vi} lines in the optical wavelength region, which are seen for example in very hot (\Teff $>$ 100\,000\,K) PG1159 stars, are from levels with high principal quantum number ($n>6$) and excitation energies above 1\,000\,000 cm$^{-1}$. We do not identify any of these lines. This suggests  a temperature for \jn that is slightly below 100\,000\,K, such that these energy states are rarely populated. 


\subsection{Spectral fitting}

As mentioned, the \ion{O}{iv} emission lines are formed in the stellar wind, whereas our static model atmosphere predicts absorption lines; see, for example, the strongest emission complex around 3400\,\AA. Our analysis relies on the ionization balance between \ion{O}{v} and \ion{O}{vi} lines, which are all forming in the photosphere. As mentioned above, we compared the relative strength of the \ion{O}{vi} doublet and the \ion{O}{v} line at 5114\,\AA, which depends on both the effective temperature and the surface gravity. To fit the observation, we fix \logg\ at a particular value and then vary \Teff. In the relevant parameter range, increasing \Teff increases the strength of the \ion{O}{vi} doublet but decreases the strength of the \ion{O}{v} line. A similar effect is observed when decreasing \logg at a fixed \Teff. At \Teff below 80\,000 and above 110\,000\,K, \ion{O}{vi} and \ion{O}{v} are too weak at any \logg. We finally adopt \Teff = 95\,000 $\pm$ 15\,000\,K and \logg = $5.5 \pm 0.5$. A look at other \ion{O}{v} lines reveals that some of them are well reproduced by the model (e.g., in the regions 4120--4220 and 5570--5610\,\AA), while others are too strong in the model (e.g., at 4400 and around 4550\,\AA) because of possible uncertainties in the model atom as discussed above. Our values for temperature and gravity are significantly different from the estimate by \citetalias{El-Badry2023} (\Teff = 130\,000\,K, \logg = 6.0). The high-temperature estimate was motivated by findings from static TMAP models of PG1159 stars, which show the \ion{C}{iv} 5801/12 doublet in emission only at such high temperatures. Indeed, our final model shown in Fig.\,\ref{fig:J0546+0836} does not show this line at all. Coming from low temperatures, at around 100\,000\,K, the line turns from absorption into emission \citep{1992LNP...401..273W}. Like the \ion{O}{iv} lines, the strong and broad \ion{C}{iv} 5801/12\,\AA\ feature is formed in the stellar wind and is therefore not matched by our static model.

We estimate the carbon abundance (by mass) to be C = 0.10 $\pm$ 0.05. At this value, the \ion{C}{iv} absorption lines at 3690\,\AA\ fits well. Other \ion{C}{iv} lines that are not detectable in observation (at 4441 and 5471\,\AA) indicate that this value cannot be larger. An upper limit for the helium abundance was found by models with different He abundances. We adopt He $<$ 0.05. A model with 5\% helium exhibits significant \ion{He}{ii} absorption lines at 4686, 5412, and 6560\,\AA, in contrast to the observation. 

\begin{figure*}
 \centering  \includegraphics[width=0.9\textwidth]{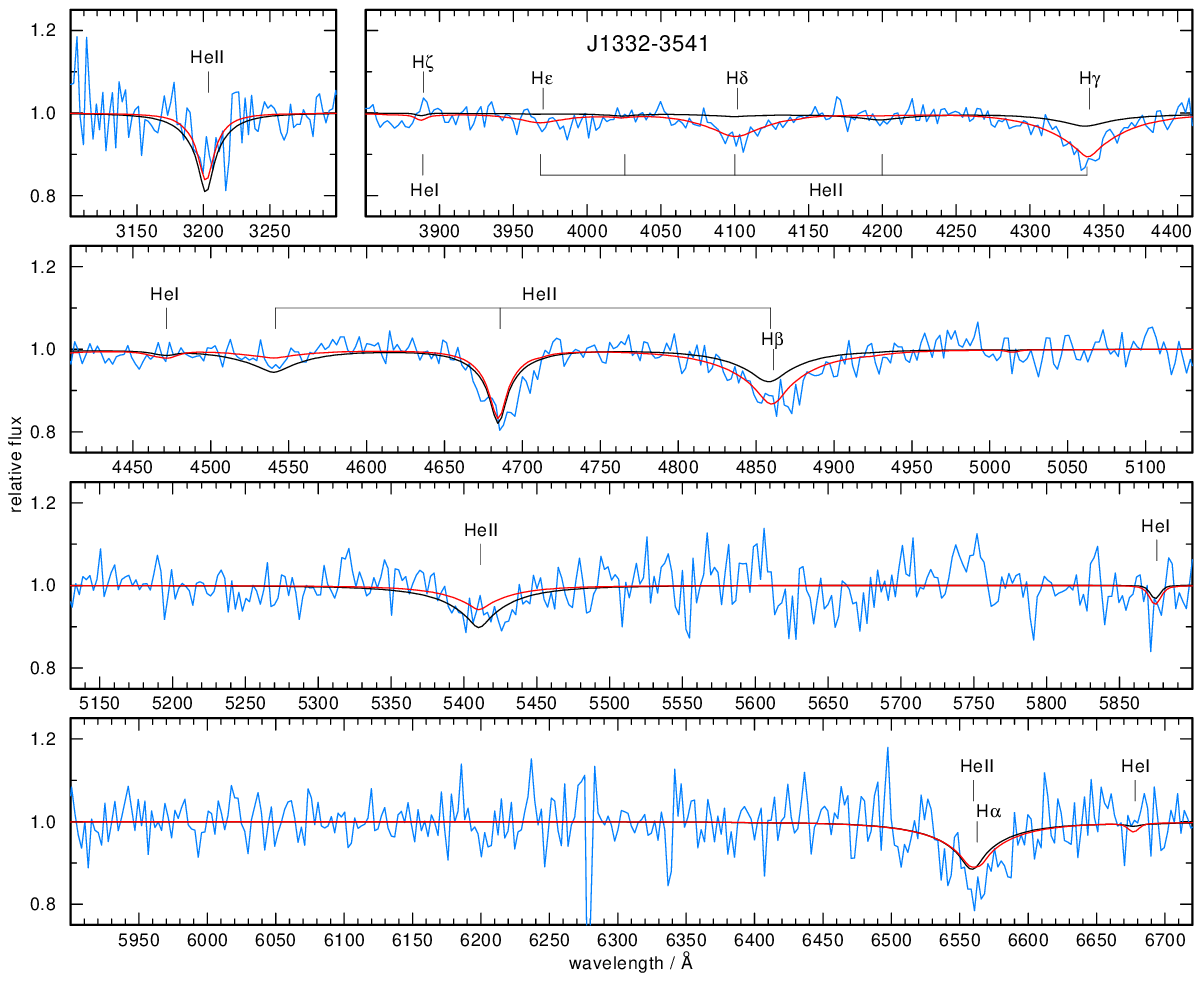}
 \caption{Keck/LRIS spectrum of the hydrogen-dominated DAO WD \jnd shifted to rest wavelength and compared to our best-fit model (red graph) with \Teff = 65\,657\,K, \logg = 8.38, H = 0.65, and He = 0.35 (mass fractions). The black graph shows the best-fit pure-helium DO model, which has \Teff = 79\,996\,K and \logg = 7.93.  }
\label{fig:J1332-3541}
\end{figure*}

\section{J0927-6335}
\label{sect:jnz}

We analyzed the spectrum of \jnz obtained by \citetalias{El-Badry2023} on MJD 59974.21 with the Magellan Echellette (MagE) spectrograph on the 6.5\,m Magellan Baade telescope at Las Campanas Observatory in Chile. The spectral resolution is $R \approx 5500$. To compare with models, we rectified the spectrum. However, this procedure was rather difficult towards the blue, at wavelengths below about 4000\,\AA, because of relatively low S/N. Therefore, the definition of the continuum is rather subjective and needs to be considered during the model fitting. We smoothed the spectrum with a Gaussian of 2\,\AA\ in width. The model spectra were folded accordingly with a Gaussian that accounts for smoothing and instrumental broadening.

The spectrum of \jnz is shown in Fig.\,\ref{fig:J0927-6335}. \citetalias{El-Badry2023} proposed that it is dominated by \ion{C}{iv} and \ion{O}{iv} lines and these authors set a lower limit to \Teff of 70\,000\,K because of the weakness of several \ion{C}{iii} lines and an upper limit to 95\,000\,K because of the absence of \ion{O}{v} lines (assuming \logg = 7.0, C = 0.72, O = 0.28). However, we find that the \ion{C}{iii} lines with the most securely known oscillator strengths are strong, and the relative strengths of the lines of the two carbon ions suggest a temperature of below 70\,000\,K. Let us first describe our line identifications in detail. 

\subsection{\ion{C}{iii} and \ion{C}{iv} lines}

A number of \ion{C}{iii} lines are detectable. Some of them are quite strong, for example the multiplet at 4070\,\AA. For many lines, oscillator strengths are unknown and line profiles cannot be calculated. We have indicated those of them that are probably visible in the observation in Fig.\,\ref{fig:J0927-6335} with the label ``\ion{C}{iii}, no model''. It might be possible that a number of unidentified lines in the spectrum stem from \ion{C}{iii}. As mentioned by \citetalias{El-Badry2023}, there are no strong lines of most metals besides C and O; but the limited S/N of the spectrum makes it difficult to judge whether other absorption features are real.

The most prominent feature in the spectrum of \jnz is the absorption trough at 4600--4750\,\AA,\ which is a defining characteristic of the PG1159 spectral class. However, there is a difference in that the red half of the trough, centered at 4685\,\AA, is predominantly caused by \ion{He}{ii} 4686\,\AA\ (transition between levels with $n=3-4$) in PG1159 stars, but in \jnz it is entirely stemming from \ion{C}{iv} $n=6-8$ transitions. This coincidence occurs because \ion{He}{ii} and \ion{C}{iv} are hydrogen-like ions and any $n-n'$ transition of \ion{He}{ii} corresponds to a $2n-2n'$ transition of \ion{C}{iv}. Other broad \ion{C}{iv} absorption features are at 3690\,\AA\ ($n=6-9$ transitions) and 5470\,\AA\ ($n=7-10$ transitions). Three other such prominent features are detectable, but these are not included in our NLTE model atom because of the lack of atomic data (labeled in Fig.\,\ref{fig:J0927-6335} with ``\ion{C}{iv}, no model''). Two of these features are located at 4230\,\AA\ ($n=7-12$) and 5865\,\AA\ ($n=8-13$), and the third is causing the broad absorption line at 6560\,\AA\ ($n=8-12$), which does not stem from \ion{He}{ii} ($n=4-6$) or H$\alpha$ ($n=2-3$).

\subsection{\ion{O}{iii} and \ion{O}{iv} lines}

Our models with temperatures of around \Teff = 60\,000\,K predict lines from \ion{O}{iii} and \ion{O}{iv}, but these are relatively weak and hard to identify in view of the relatively poor S/N of the observed spectrum; see, for example, the \ion{O}{iii} lines at 3962, 5268, 5508, and 5592\,\AA\ in Fig.\ref{fig:J0927-6335}, or the \ion{O}{iv} line at 4344\,\AA. The strongest oxygen lines predicted by our models are in the range 3700--3800\,\AA. As mentioned, the practical problem here is that the S/N of the observed spectrum is so poor that it is very difficult to recognize the continuum during the normalization procedure. It could well be that the deep absorption feature of  approximately
20\,\AA\ in width at 3760\,\AA\  is caused by three strong and blended \ion{O}{iii} lines.

\begin{table*}[t]
\begin{center}
\caption{Atmospheric properties and other parameters of our program stars.
\tablefootmark{a} 
}
\label{tab:resultsall}
\begin{tabular}{cccc}
\hline 
\hline 
\noalign{\smallskip}
Parameter                 & \jn                    & \jnz                      & \jnd              \\
\hline
\noalign{\smallskip}
Spectral type             & [WC]--PG1159           & PG1159                    & DAO               \\
\noalign{\smallskip}
\Teff (K)                 & $95\,000 \pm 15\,000$  & $60\,000 \pm 5000$        & $65\,657 \pm 2390$\\
$\logg$ (cm\,s$^{-2}$) &  $5.5 \pm 0.5$      & $7.0 \pm 0.5$             & $8.38\pm0.08$     \\
\noalign{\smallskip}
H                         & $-$                    & $-$                       & $0.65 \pm 0.04$   \\   
He                        & $<0.05$                & $0.05 \pm 0.05$           & $0.35 \pm 0.04$   \\   
C                         & $0.10 \pm 0.05$        & $0.47 \pm 0.25$           & $-$               \\ 
O                         & $0.90 \pm 0.05$        & $0.48 \pm 0.25$           & $-$               \\ 
\noalign{\smallskip}
$R$ (\Rsol)               & $0.21^{+0.06}_{-0.11}$ & $0.030^{+0.012}_{-0.010}$ & $0.010\pm 0.001$  \\
\noalign{\smallskip}
$M$ (\Msol)               & $0.52^{+0.31}_{-0.04}$ & $0.32^{+0.16}_{-0.12}$    & $0.89 \pm 0.03$   \\
\noalign{\smallskip}
$L$ (\Lsol)               & $3218^{+6345}_{-2851}$ & $21.1^{+19.5}_{-16.1}$    & $1.7^{+0.6}_{-0.5}$    \\
\noalign{\smallskip}
$E(B-V)$ (mag) (SED fit)  & 0.29                   & 0.15                      & 0.032                  \\
\noalign{\smallskip}
$d$ (kpc) (SED fit)       & $15^{+4}_{-8}$         & $2.5\pm0.9$               & $1.09^{+0.12}_{-0.11}$ \\
\noalign{\smallskip}
$d$ (kpc) ({\it Gaia})    & $4.0^{+2.3}_{-1.7}$    & $4.5^{+2.2}_{-1.7}$       & $1.63^{+1.19}_{-0.68}$ \\
\noalign{\smallskip}
\hline
\end{tabular} 
\tablefoot{  
\tablefoottext{a}{Element abundances given in mass
    fractions. Stellar radius, mass, and luminosity for \jn derived from post-AGB tracks of \cite{2006A&A...454..845M}, for \jnz from cooling tracks of \cite{Bedard+2020}, and for \jnd from cooling tracks of \cite{Renedo+2010}, as displayed in
    Fig.\,\ref{fig:gteff}. The {\it Gaia} distances were determined by
     \citetalias{El-Badry2023} assuming a flat prior on $M_{G,0}$. }  } 
\end{center}
\end{table*}

\begin{figure*}
 \centering  \includegraphics[width=0.8\textwidth]{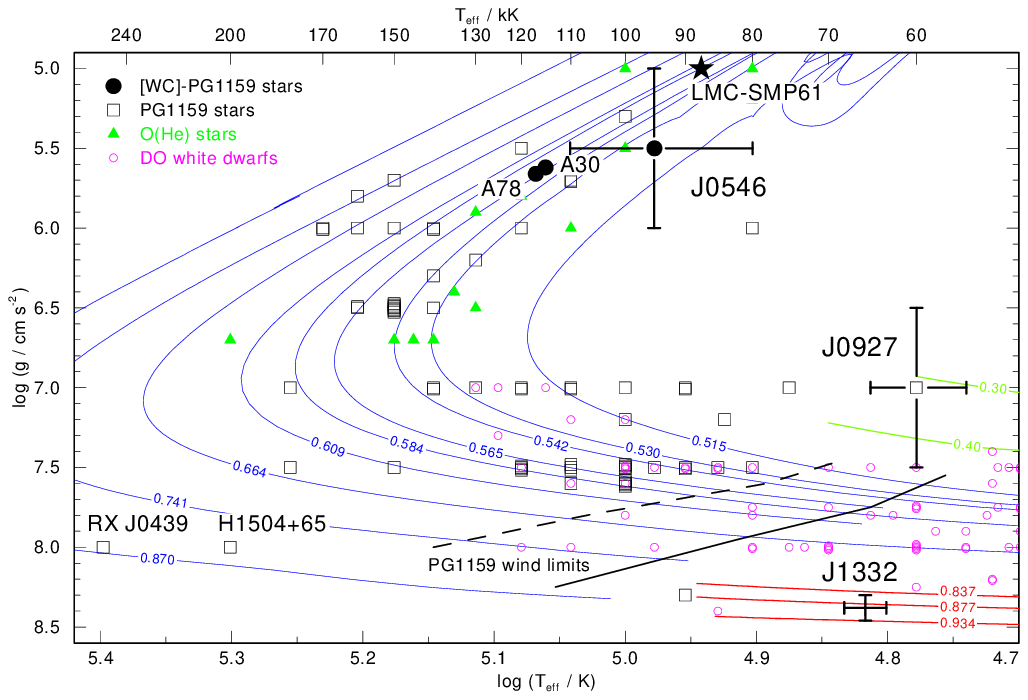}
 \caption{Positions of the two C-O dominated stars \jn and \jnz as well as the DAO WD \jnd in the Kiel diagram, 
      together with other hydrogen-deficient stars: PG1159 stars, O(He) stars, and DO WDs. A30 and A78 are [WC]--PG1159 transition objects. H1504$+$65 and RX\,J0439 are PG1159 stars that are He-deficient. LMC-SMP\,61 is the [WC] central star discussed in the text. We show three sets of evolutionary tracks that could be representative of our program stars, labelled with the stellar mass in solar units: VLTP post-AGB stars 
      \citep[blue, ][]{2006A&A...454..845M} and WD coolings tracks of \cite{Bedard+2020} (green) and \cite{Renedo+2010} (red).
      The black line indicates the PG1159 wind limit
  according to \citet{2000A&A...359.1042U}, meaning that the mass-loss
  rate of the radiation driven wind at this position of the evolutionary tracks
  becomes so weak that gravitational settling of heavy elements
  is able to remove them from the atmosphere. Thus, no PG1159
  stars should be found at significantly cooler temperatures. The dashed line is the
  wind limit assuming a ten-times-lower mass-loss rate.
      }
\label{fig:gteff}
\end{figure*}

\subsection{Spectral fitting}

We started our analysis by calculating a model with the parameters estimated by \citetalias{El-Badry2023}: \Teff = 80\,000\,K, \logg = 7, C = 0.72, O = 0.28. We found the temperature to be significantly lower, namely of around 60\,000\,K, because otherwise most of the \ion{C}{iii} lines are too weak in the model. \citetalias{El-Badry2023} disfavored a temperature of below 70\,000\,K because their spectral models would predict excessively strong  \ion{C}{iii} lines in the spectral region 5050--5300\,\AA. In particular, a \ion{C}{iii} line at 5130\,\AA\ is too strong in their 65\,000\,K model. However, this line must be regarded as uncertain, because there are no oscillator strengths listed in the NIST and Kentucky\footnote{\url{https://linelist.pa.uky.edu/atomic/}} databases. This is probably because this \ion{C}{iii} line  is a blend of $5g-7h$ transitions in both the singlet and triplet systems, and is therefore difficult to measure in the laboratory. Consequently, this line feature is not included in our models. We emphasize that the strongest \ion{C}{iii} lines are located at shortest wavelengths ---for example, at 3886 and 4070\,\AA--- and require, as mentioned, a lower temperature, that is, near 60\,000\,K. Reducing the temperature to this value also causes an  increase in the strength of the \ion{C}{iv} lines, albeit less drastic. It is therefore possible to decrease the C/O ratio in order to increase the strengths of the oxygen lines in the 3700--3800\,\AA\ region discussed above such that we get a better overall fit. We finally adopted the following parameters: \Teff = 60\,000 $\pm$ 5000\,K, \logg = $7.0 \pm 0.5$, and equal amounts of C and O by mass. The error limits for the surface gravity were estimated from models with \logg = 6.5 and 7.5. The wings of the strongest \ion{C}{iv} lines in the 4600--4750\,\AA\ region and at 5471\,\AA\ are too narrow and too wide compared to the observation, respectively. 

An upper limit for the helium abundance of He $<$ 0.1 was determined by \citetalias{El-Badry2023}. Traces of helium can perhaps be identified by the appearance of the \ion{He}{ii} 5412\,\AA\ line. At He = 0.05, we note a weak absorption feature in the model that seems to match the observation. \ion{He}{ii} 6560\,\AA\ is stronger in the model, but the strong \ion{C}{iv} line dominates in the observation. The  \ion{He}{ii} 4859\,\AA\  absorption line is weaker in the model. The contribution from \ion{He}{ii} 4686\,\AA\ to the absorption trough is negligible and \ion{C}{iv} dominates. We regard the detection of helium as uncertain and better observations are necessary in order to confirm the detection. We finally adopt the composition He = $0.05 \pm 0.05$, C = $0.47 \pm 0.25$, and O = $0.48 \pm 0.25$. 

\section{J1332$-$3541}
\label{sect:jnd}

\citetalias{El-Badry2023} obtained a spectrum for \jnd on MJD 60054.52 with LRIS on the 10\,m Keck-I telescope on Maunakea, Hawaii. The spectral resolution is $R \approx 1000$. These latter authors classified the star as a DO WD and estimated \Teff $\approx$ 70\,000\,K and \logg $\approx$ 7.5.

As a first attempt, we used the pure-He model grid computed by \cite{Reindl+2023} and performed a $\chi^2$ fit to the LRIS spectrum. We found \Teff = 79\,996 $\pm$ 4215\,K and \loggw{7.93\pm0.09}. However, the fit did not reproduce the observation well (black graph in Fig.\,\ref{fig:J1332-3541}). In particular, several of the \ion{He}{ii} lines were too weak in the best-fit model compared to the observation.  

We remind the reader that the \ion{He}{ii} Pickering series has lines almost exactly coincident with the  \ion{H}{i} Balmer series. This makes it difficult to determine whether or not a hot star with strong \ion{He}{ii} lines also contains H  by visual inspection alone. \citetalias{El-Badry2023} attributed all the lines in the spectrum of  \jnd to \ion{He}{ii}, but here we consider the alternative possibility that hydrogen is present in the atmosphere of this star. To investigate this possibility, we repeated the fit with a grid by \cite{Reindl+2023} that considers opacities from H and He. We derived \Teff = 65\,657 $\pm$ 2390\,K, \loggw{8.38\pm0.08}, H = $0.65\pm0.04$, and He = 0.35 (mass fractions). With these parameters, most of the spectrum of \jnd is well reproduced, and in particular, the regions containing the Balmer lines are significantly better fit than in the pure-He model. However, \ion{He}{ii} 4686, 5412,  and 6560\,\AA\ are slightly broader in the observation than in the model (red line in Fig.\,\ref{fig:J1332-3541}). A higher-S/N spectrum could help us to investigate the possible presence of a mild version of the \ion{He}{ii} line problem \citep{Werner1995}. If indeed present,  the derived atmospheric parameters could suffer from systematic uncertainties. We note that the origin of the \ion{He}{ii} line problem remains elusive, and therefore it is not possible to obtain a robust estimate of the systematic errors.

The models of \cite{Reindl+2023} assume a homogeneous distribution of H and He in the atmosphere. However, some DAO WDs are known to have a chemically stratified atmosphere, with a very thin H-rich layer on top of the He-rich envelope. These objects are thought to be DOs transforming into DAs as residual H mixed into the envelope gradually floats up to the surface \citep{2016ApJ...833..127M,Bedard+2020,2023ApJ...946...24B}. To test the possibility that \jnd has a stratified atmosphere, we attempted a spectral fit using the stratified model grid of \cite{Bedard+2020}. We obtained a poorer fit (not shown here), indicating that \jnd likely has a homogeneous atmosphere.

\begin{figure}
 \centering  \includegraphics[width=\linewidth]{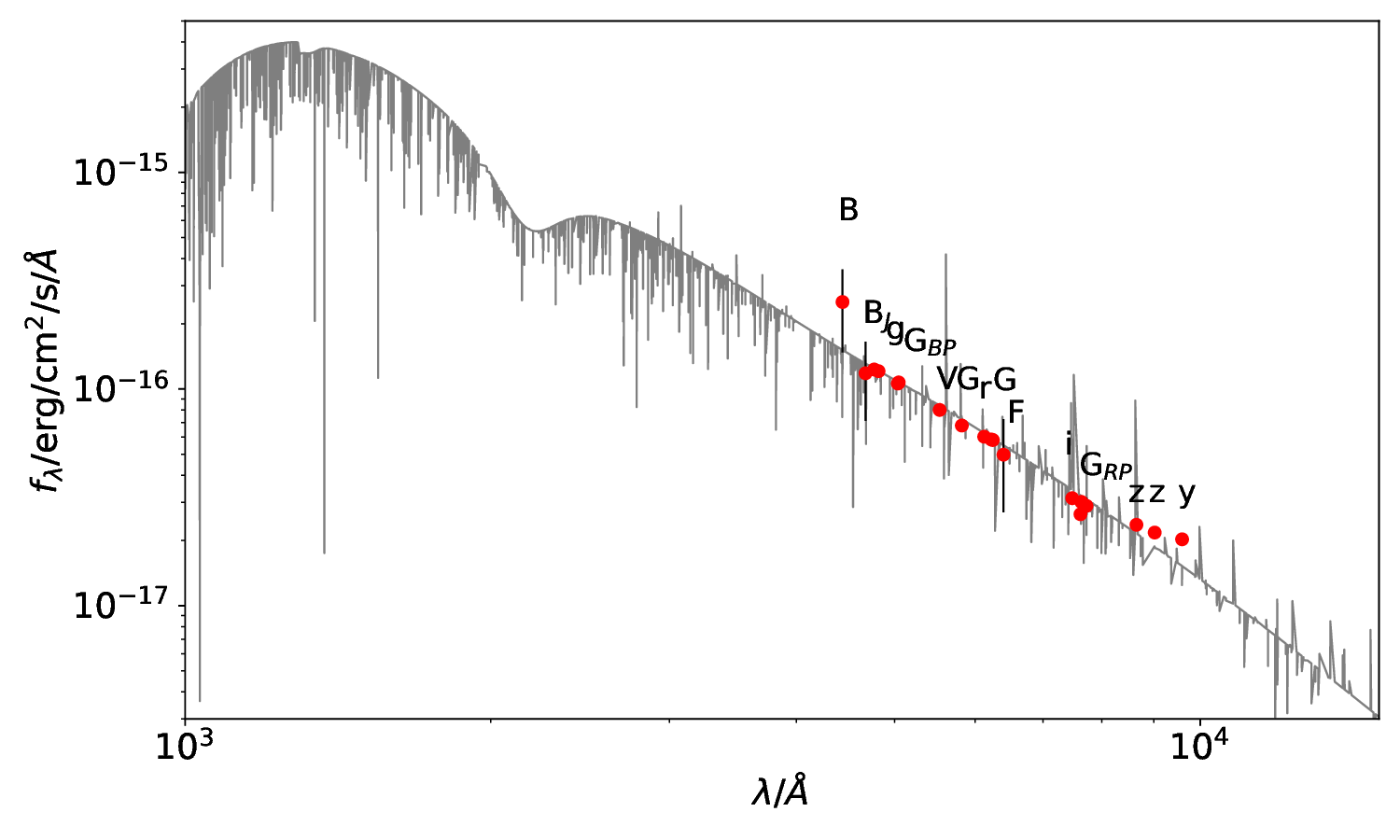}
 \centering  \includegraphics[width=\linewidth]{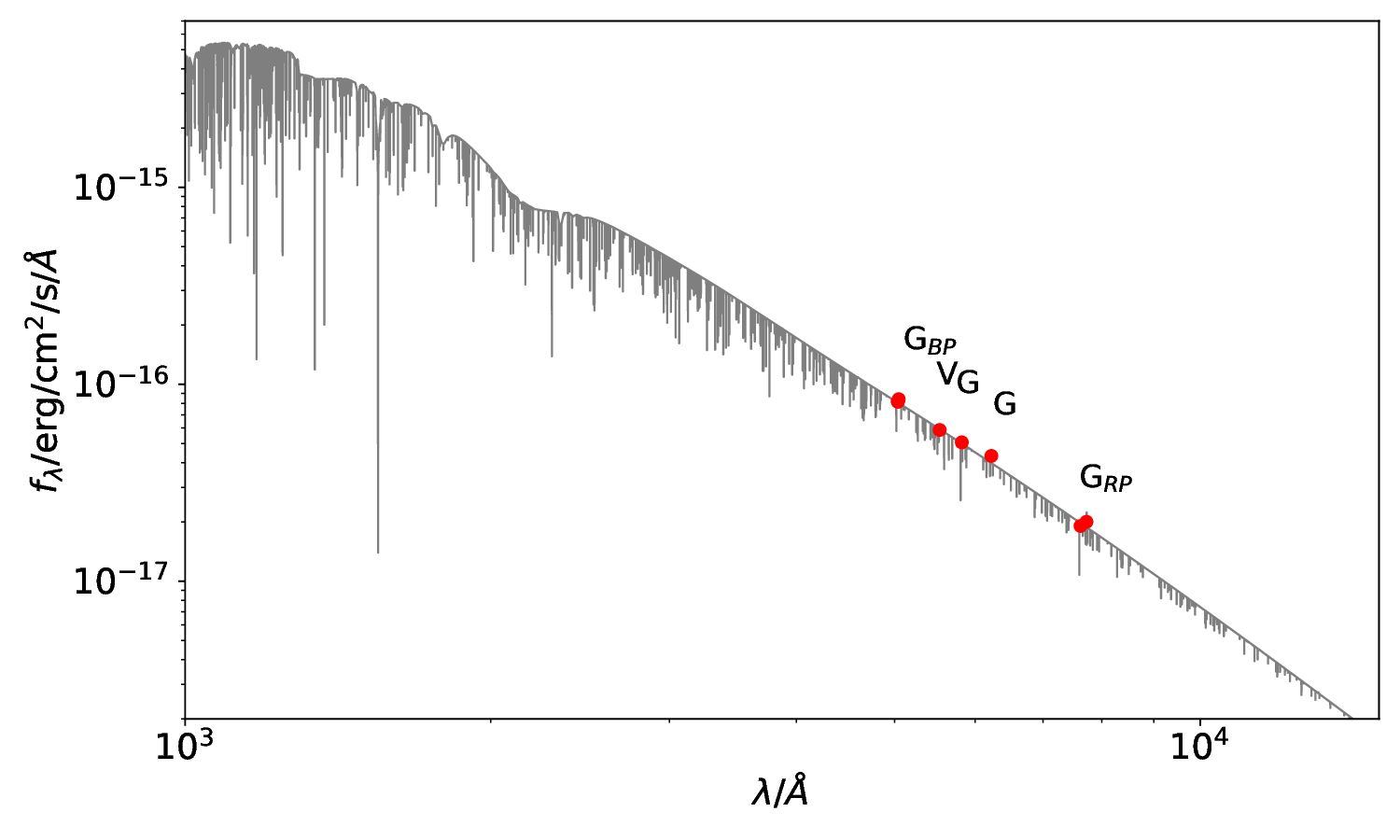}
 \centering  \includegraphics[width=\linewidth]{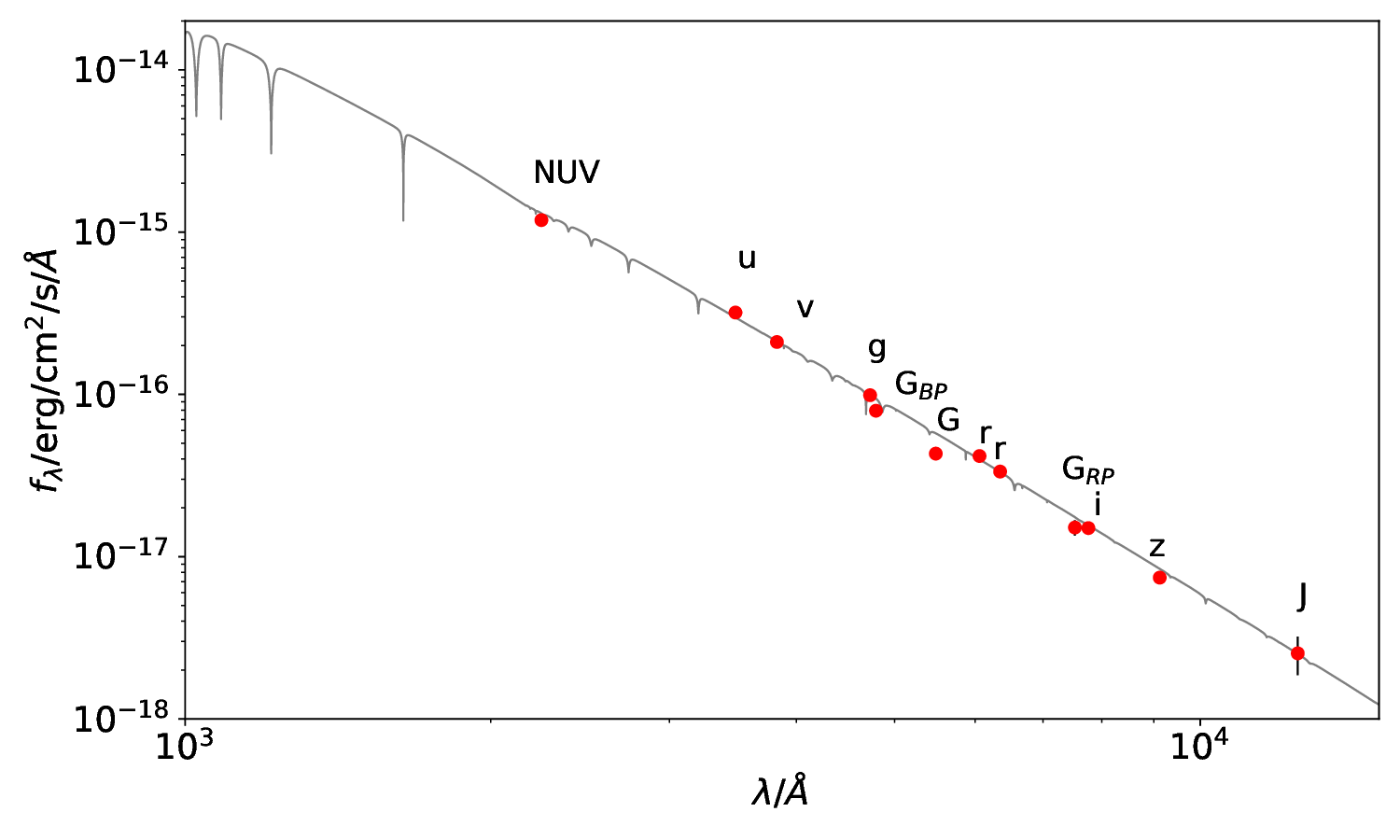}
 \caption{SED fits to \jn (top panel), \jnz (middle), and  \jnd (bottom).  }
\label{fig:sed}
\end{figure}

\section{Masses, radii, luminosities, and spectroscopic distances}
\label{sect:sed}

Luminosities, radii, and masses are determined via interpolation from evolutionary tracks. For the DAO WD, we employed evolutionary tracks for CO-core WDs from \cite{Renedo+2010}; for \jn, we employed the tracks from \cite{2006A&A...454..845M}; and for \jnz, we used theoretical cooling sequences for He-rich atmosphere CO-core WDs of the Montreal White Dwarf Group\footnote{\url{https://www.astro.umontreal.ca/~bergeron/CoolingModels/}} \citep{Bedard+2020}. The results are valid provided that such tracks can safely be applied to these stars.

Figure\,\ref{fig:gteff} shows the location of our program stars in the Kiel ($g$ -- \Teff) diagram together with other H-deficient objects, including the PG1159 stars\footnote{According to an unpublished list based on \cite{2006PASP..118..183W} and maintained by one of the authors (KW).}, the helium-dominated O(He) stars \citep[e.g.,][]{2014A&A...566A.116R,2023MNRAS.519.2321J}, and DO WDs \citep[e.g.,][]{1996A&A...314..217D,2023MNRAS.519.2321J}. Also shown are the locations of the two [WC]--PG1159 transition objects Abell~30 and Abell~78 as well as the [WC] star LMC-SMP\,61 introduced in Sect.\,\ref{sect:jn}.


Subsequently, we collected available photometry from various catalogs in order to estimate the spectrophotometric distances of our stars via fits of our best-fitting models to their observed SED. To this end, we relied on the results from our spectral analysis and assumed the radius to be fixed within the determined uncertainties, while the distance and interstellar extinction were considered as free parameters. 

For \jnd, we employed the $\chi ^2$ SED fitting routine described in \cite{Heber+2018} and \cite{Irrgang+2021} to find the global best fit using the model grid of \cite{Reindl+2023}. The SED fit is shown in the bottom
panel of Fig.\,\ref{fig:sed}, and we derive $E(B-V)=0.038$\,mag and a spectroscopic distance of $1.09^{+0.12}_{-0.11}$\,kpc. For \jn and \jnz, we used the best-fitting models computed in this work and performed the fit by eye. For \jn, we find that the SED is best reproduced with $E(B-V)=0.29$\,mag (as suggested by the \cite{2022A&A...661A.147L} map) and obtain a spectroscopic distance of $15^{+4}_{-8}$\,kpc. The SED fit of this star is shown in the top panel of Fig.\,\ref{fig:sed}. With $E(B-V)=0.15$\,mag, which was also adopted by \cite{El-Badry2023}, we find that the SED of \jnz is well reproduced at a spectroscopic distance of $2.5\pm0.9$\,kpc (middle panel of Fig.\,\ref{fig:sed}).

Within the error limits, our spectroscopic distances are similar to the \emph{Gaia} distances derived by \citetalias{El-Badry2023} assuming a flat prior on the absolute magnitude $M_{G,0}$ in the \emph{Gaia} bandpass (Table\,\ref{tab:resultsall}), except for \jn.  Our value of 15\,kpc for this latter source is about four times larger than the \emph{Gaia} distance. The large distance reflects the low surface gravity, hence the high luminosity determined by our spectral analysis. Given the proper motion of \jn\ of $\mu = 76.1\,\rm mas\,yr^{-1}$, a distance of 15 kpc would correspond to a rather large tangential velocity of $\rm 5400\,\rm km\,s^{-1}$. Such a large velocity would be difficult to explain even in the D$^6$ scenario, suggesting that our $\log g$ may be underestimated somewhat, or that the star was produced through some other mechanism. 

\section{Summary and discussion}
\label{sect:summary}

We performed detailed line identifications and NLTE spectral analyses of the three hottest hypervelocity stars discovered spectroscopically and characterized preliminarily by \citetalias{El-Badry2023}. Below we summarize our main findings (see also Table\,\ref{tab:resultsall}) and compare them with this latter work. We then comment further on the individual objects. 

(i) We confirm that \jn and \jnz have carbon- and oxygen-dominated atmospheres. \jnz might also have minute amounts of helium. Both stars are significantly cooler than estimated by \citetalias{El-Badry2023}, namely we find 95\,000 and 60\,000\,K, respectively, instead of 130\,000 and 80\,000\,K as found by \citetalias{El-Badry2023}. Temperature and gravity (\logg = 5.5 and 7.0, respectively) place the stars in a similar location in the Kiel diagram to low-mass ($\approx 0.5$\,\Msol) evolutionary tracks of hydrogen-deficient post-AGB objects that suffered a late helium-shell flash (Fig.\,\ref{fig:gteff}). 

(ii) We confirm that \jnd is a hot WD, but we find that it is not a helium-rich DO star as originally thought, but a hydrogen-rich DAO star with slightly supersolar helium abundance. We infer a relatively high mass of 0.89\,\Msol.

\paragraph{\jn}
Spectroscopically, this is a [WC]--PG1159 transition-type object, meaning that it exhibits both emission lines forming in a wind and absorption lines forming in the photosphere. Two other stars with this property are known, namely the central stars of the planetary nebulae Abell~30 and Abell~78. These stars are slightly hotter versions of \jn (Fig.\,\ref{fig:gteff}) and we argue that \jn should have a similar mass-loss rate (about \Mdot = $10^{-8}$ \Msol yr$^{-1}$). 

However, the main difference between \jn and the PG1159 class is the helium abundance. With two notable exceptions, PG1159 stars have large amounts of helium (at least about 30\%). The prototype PG1159$-$035 for instance has He = 0.33, C = 0.48, O = 0.17, and Ne = 0.02 \citep{2004A&A...427..685W}. \jn on the other hand is devoid of helium and is composed of C = 0.1 and O = 0.9. Among the currently known 68 PG1159 stars, there are two members that were also found to lack helium: H1504$+$65 (C = 0.46, O = 0.46, Ne = 0.06, Mg = 0.02) and RX\,J0439.8$-$6809 (C = 0.50, O = 0.49, Ne = 0.01) \citep{2015A&A...584A..19W}. In this latter work, we discuss the possible origins of this composition (extraordinary strong mass loss after a He-shell flash or binary CO--WD merger). In particular, for RX\,J0439.8$-$6809, we considered a supernova runaway because the star has a high radial velocity ($+$220\,km\,s$^{-1}$) and is located in the Galactic halo at a distance of 9.2\,kpc towards the Large Magellanic Cloud. Unfortunately, the star is too faint (V = 21.74) for \emph{Gaia} proper-motion measurements.  

H1504$+$65 and RX\,J0439.8$-$6809 are the hottest WDs known and they are at the hot end of the WD cooling sequence (Fig.\,\ref{fig:gteff}). However, \jn is located at post-AGB tracks {before} their high-temperature knee. This means that \jn\ ---if these tracks are indeed representative of this object--- is performing He-shell burning. This is difficult to reconcile with its helium deficiency on the surface unless one assumes that it has a helium layer below the C-O envelope that was accreted by a helium-rich WD at the explosion of the companion.

\cite{2019ApJ...872...29Z} computed long-term evolutionary models with Modules for Experiments in Stellar Astrophysics (MESA) for SN Iax postgenitors (the accretors) and noted that these models and those of runaway companions (the donors) may look very much alike. Particularly interesting in our context are their ``abnormal cases'', where the envelope entropy is high and as a consequence the simulations fail to reach the cooling track for numerical reasons. One such model expands to AGB-star dimensions with super-Eddington luminosity (with \Teff $\approx$ 3500\,K and $L=10^4$\,\Lsol). The authors noted that wind mass loss should be included in future simulations of these phases. The wind will probably also suppress the radiative levitation of iron and nickel, an effect shown by their models in the hot luminous phases. We note again that \jn indeed exhibits a radiation-driven wind. It may be reasonable to assume that respective evolutionary tracks of stars descending from the AGB region should be similar to the post-AGB tracks shown in our Fig.\,\ref{fig:gteff}. 

\cite{2019ApJ...872...29Z} emphasize the role of gravitational settling in the appearance of the SN Iax postgenitors. Figure\,\ref{fig:gteff} shows two wind limits that denote the line in the Kiel diagram where PG1159 stars turn into helium-rich DO WDs. When approaching this limit, the radiation-driven wind becomes so weak that it is no longer able to counteract gravitational settling of carbon and oxygen. \jn is very luminous and this effect is therefore unimportant, which means that the abundances during the postgenitor evolution are unchanged. However, this is not the case for the other two stars that we analyzed.

\paragraph{\jnz}

Spectroscopically this is a PG1159 star. However, like \jn, it has a lower helium abundance ($\rm He \lesssim 0.05$ or even zero) than the other stars of this class (with the exception of the two He-deficient objects discussed above). It is also cooler than any PG1159 star and close to the wind limit in the Kiel diagram (Fig.\,\ref{fig:gteff}). The possible minute amount of helium could signal that the star is about to transform into a DO WD with helium diffusing to the surface. It is therefore possible that the  abundances of the original postgenitor have changed and that even heavier elements than carbon and oxygen are depleted by gravitational settling.

\paragraph{\jnd}

Surprisingly, this star is not a hot helium-rich DO WD, as originally proposed, but a hydrogen-dominated DAO WD with a roughly solar helium abundance. We find that the atmosphere is chemically homogeneous, which would be a sign that this is not a former DO WD that is transforming into a DA. 
However, it is still possible that \jnd is a DO-to-DA transition object. \cite{2023ApJ...946...24B} performed detailed simulations of the H float-up process and showed that this phenomenon is governed by an interplay between atomic diffusion (which tends to push H upward) and the radiative wind (which tends to homogenize the outer layers). These latter authors also showed that the time required to achieve the DO-to-DA transformation depends mainly on the amount of residual H. If the initial H abundance is relatively low (e.g., the $\log$ H = $-4.5$ case shown in their fig.\,3), it takes more time for the star to develop a H-rich surface. During this time, the WD cools and its wind consequently fades, thereby allowing diffusion to produce a stratified atmosphere. However, if the initial H abundance is higher (e.g., the $\log$ H = $-3.0$ case shown in their fig.\,2), the DO-to-DA transition takes place earlier, when the wind is still powerful enough to homogenize the outer envelope. This scenario indeed produces a DAO WD with a homogeneous atmosphere, as observed for \jnd.

The presence of hydrogen is puzzling and difficult to reconcile with the D$^6$ scenario. It is possible that the relatively massive WD collected hydrogen during its fast cruise through the interstellar medium. Only small amounts of hydrogen are needed for a thin, optically thick mantle. Any accreted metals rapidly sink into the interior.

Our study confirms the D$^6$ scenario is a plausible hypothesis for the existence of the carbon- and oxygen-dominated stars \jn and \jnz. Ultraviolet spectroscopy is needed to search for other metals. For the third object, the hydrogen-dominated DAO WD \jnd, the D$^6$ scenario has difficulty in explaining the atmospheric composition.

\begin{acknowledgements} 
NR is supported by the Deutsche Forschungsgemeinschaft (DFG) through grant RE3915/2-1. The TMAD tool (\url{http://astro.uni-tuebingen.de/~TMAD}) used for this paper was constructed as part of the activities of the German Astrophysical Virtual Observatory. This research has made use of NASA's Astrophysics Data System and the SIMBAD database, operated at CDS, Strasbourg, France. This research has made use of the VizieR catalogue access tool, CDS, Strasbourg, France. This work has made use of data from the European Space Agency (ESA) mission {\it Gaia}.
\end{acknowledgements}

\bibliographystyle{aa}
\bibliography{aa}

\begin{thebibliography}{45}
\expandafter\ifx\csname natexlab\endcsname\relax\def\natexlab#1{#1}\fi

\bibitem[{{Bauer} {et~al.}(2021){Bauer}, {Chandra}, {Shen}, \&
  {Hermes}}]{Bauer2021}
{Bauer}, E.~B., {Chandra}, V., {Shen}, K.~J., \& {Hermes}, J.~J. 2021, \apjl,
  923, L34

\bibitem[{{B{\'e}dard} {et~al.}(2023){B{\'e}dard}, {Bergeron}, \&
  {Brassard}}]{2023ApJ...946...24B}
{B{\'e}dard}, A., {Bergeron}, P., \& {Brassard}, P. 2023, \apj, 946, 24

\bibitem[{{B{\'e}dard} {et~al.}(2020){B{\'e}dard}, {Bergeron}, {Brassard}, \&
  {Fontaine}}]{Bedard+2020}
{B{\'e}dard}, A., {Bergeron}, P., {Brassard}, P., \& {Fontaine}, G. 2020, \apj,
  901, 93

\bibitem[{{Burmester} {et~al.}(2023){Burmester}, {Ferrario}, {Pakmor},
  {Seitenzahl}, {Ruiter}, \& {Hole}}]{Burmester2023}
{Burmester}, U.~P., {Ferrario}, L., {Pakmor}, R., {et~al.} 2023, \mnras
  [\eprint[arXiv]{2305.05192}]

\bibitem[{{Chandra} {et~al.}(2022){Chandra}, {Hwang}, {Zakamska}, {Blouin},
  {Swan}, {Marsh}, {Shen}, {G{\"a}nsicke}, {Hermes}, {Putterman}, {Bauer},
  {Petrosky}, {Dhillon}, {Littlefair}, \& {Ashley}}]{Chandra2022}
{Chandra}, V., {Hwang}, H.-C., {Zakamska}, N.~L., {et~al.} 2022, \mnras, 512,
  6122

\bibitem[{{Dreizler} \& {Werner}(1996)}]{1996A&A...314..217D}
{Dreizler}, S. \& {Werner}, K. 1996, \aap, 314, 217

\bibitem[{{El-Badry} {et~al.}(2023){El-Badry}, {Shen}, {Chandra}, {Bauer},
  {Fuller}, {Strader}, {Chomiuk}, {Naidu}, {Caiazzo}, {Rodriguez}, {Nagarajan},
  {Yamaguchi}, {Vanderbosch}, {Roulston}, {G{\"a}nsicke}, {Han}, {Burdge},
  {Filippenko}, {Brink}, \& {Zheng}}]{El-Badry2023}
{El-Badry}, K., {Shen}, K.~J., {Chandra}, V., {et~al.} 2023, The Open Journal
  of Astrophysics, 6, 28

\bibitem[{{G{\"a}nsicke} {et~al.}(2020){G{\"a}nsicke}, {Koester}, {Raddi},
  {Toloza}, \& {Kepler}}]{Gansicke2020}
{G{\"a}nsicke}, B.~T., {Koester}, D., {Raddi}, R., {Toloza}, O., \& {Kepler},
  S.~O. 2020, \mnras, 496, 4079

\bibitem[{{Guillochon} {et~al.}(2010){Guillochon}, {Dan}, {Ramirez-Ruiz}, \&
  {Rosswog}}]{Guillochon2010}
{Guillochon}, J., {Dan}, M., {Ramirez-Ruiz}, E., \& {Rosswog}, S. 2010, \apjl,
  709, L64

\bibitem[{{Heber} {et~al.}(2018){Heber}, {Irrgang}, \&
  {Schaffenroth}}]{Heber+2018}
{Heber}, U., {Irrgang}, A., \& {Schaffenroth}, J. 2018, Open Astronomy, 27, 35

\bibitem[{{Irrgang} {et~al.}(2021){Irrgang}, {Geier}, {Heber}, {Kupfer},
  {El-Badry}, \& {Bloemen}}]{Irrgang+2021}
{Irrgang}, A., {Geier}, S., {Heber}, U., {et~al.} 2021, \aap, 650, A102

\bibitem[{{Jeffery} {et~al.}(2023){Jeffery}, {Werner}, {Kilkenny}, {Miszalski},
  {Monageng}, \& {Snowdon}}]{2023MNRAS.519.2321J}
{Jeffery}, C.~S., {Werner}, K., {Kilkenny}, D., {et~al.} 2023, \mnras, 519,
  2321

\bibitem[{{Jones} {et~al.}(2019){Jones}, {R{\"o}pke}, {Fryer}, {Ruiter},
  {Seitenzahl}, {Nittler}, {Ohlmann}, {Reifarth}, {Pignatari}, \&
  {Belczynski}}]{Jones2019}
{Jones}, S., {R{\"o}pke}, F.~K., {Fryer}, C., {et~al.} 2019, \aap, 622, A74

\bibitem[{{Kurucz}(1970)}]{Kurucz1970SAOSR}
{Kurucz}, R.~L. 1970, SAO Special Report, 309

\bibitem[{Kurucz(1979)}]{kurucz_model_1979}
Kurucz, R.~L. 1979, The Astrophysical Journal Supplement Series, 40, 1, aDS
  Bibcode: 1979ApJS...40....1K

\bibitem[{{Kurucz}(1992)}]{Kurucz1992}
{Kurucz}, R.~L. 1992, in The Stellar Populations of Galaxies, ed. B.~{Barbuy}
  \& A.~{Renzini}, Vol. 149, 225

\bibitem[{{Lallement} {et~al.}(2022){Lallement}, {Vergely}, {Babusiaux}, \&
  {Cox}}]{2022A&A...661A.147L}
{Lallement}, R., {Vergely}, J.~L., {Babusiaux}, C., \& {Cox}, N.~L.~J. 2022,
  \aap, 661, A147

\bibitem[{{Livne}(1990)}]{Livne1990}
{Livne}, E. 1990, \apjl, 354, L53

\bibitem[{{Manseau} {et~al.}(2016){Manseau}, {Bergeron}, \&
  {Green}}]{2016ApJ...833..127M}
{Manseau}, P.~M., {Bergeron}, P., \& {Green}, E.~M. 2016, \apj, 833, 127

\bibitem[{{Miller Bertolami} \& {Althaus}(2006)}]{2006A&A...454..845M}
{Miller Bertolami}, M.~M. \& {Althaus}, L.~G. 2006, \aap, 454, 845

\bibitem[{{Pakmor} {et~al.}(2022){Pakmor}, {Callan}, {Collins}, {de Mink},
  {Holas}, {Kerzendorf}, {Kromer}, {Neunteufel}, {O'Brien}, {R{\"o}pke},
  {Ruiter}, {Seitenzahl}, {Shingles}, {Sim}, \& {Taubenberger}}]{Pakmor2022}
{Pakmor}, R., {Callan}, F.~P., {Collins}, C.~E., {et~al.} 2022, \mnras, 517,
  5260

\bibitem[{{Papish} {et~al.}(2015){Papish}, {Soker}, {Garc{\'\i}a-Berro}, \&
  {Aznar-Sigu{\'a}n}}]{Papish2015}
{Papish}, O., {Soker}, N., {Garc{\'\i}a-Berro}, E., \& {Aznar-Sigu{\'a}n}, G.
  2015, \mnras, 449, 942

\bibitem[{{Raddi} {et~al.}(2019){Raddi}, {Hollands}, {Koester}, {Hermes},
  {G{\"a}nsicke}, {Heber}, {Shen}, {Townsley}, {Pala}, {Reding}, {Toloza},
  {Pelisoli}, {Geier}, {Gentile Fusillo}, {Munari}, \& {Strader}}]{Raddi2019}
{Raddi}, R., {Hollands}, M.~A., {Koester}, D., {et~al.} 2019, \mnras, 489, 1489

\bibitem[{{Reindl} {et~al.}(2023){Reindl}, {Islami}, {Werner}, {Kepler},
  {Pritzkuleit}, {Dawson}, {Dorsch}, {Istrate}, {Pelisoli}, {Geier}, {Uzundag},
  {Provencal}, \& {Justham}}]{Reindl+2023}
{Reindl}, N., {Islami}, R., {Werner}, K., {et~al.} 2023, \aap, 677, A29

\bibitem[{{Reindl} {et~al.}(2014){Reindl}, {Rauch}, {Werner}, {Kruk}, \&
  {Todt}}]{2014A&A...566A.116R}
{Reindl}, N., {Rauch}, T., {Werner}, K., {Kruk}, J.~W., \& {Todt}, H. 2014,
  \aap, 566, A116

\bibitem[{{Renedo} {et~al.}(2010){Renedo}, {Althaus}, {Miller Bertolami},
  {Romero}, {C{\'o}rsico}, {Rohrmann}, \& {Garc{\'\i}a-Berro}}]{Renedo+2010}
{Renedo}, I., {Althaus}, L.~G., {Miller Bertolami}, M.~M., {et~al.} 2010, \apj,
  717, 183

\bibitem[{{Shen} {et~al.}(2018{\natexlab{a}}){Shen}, {Boubert}, {G{\"a}nsicke},
  {Jha}, {Andrews}, {Chomiuk}, {Foley}, {Fraser}, {Gromadzki}, {Guillochon},
  {Kotze}, {Maguire}, {Siebert}, {Smith}, {Strader}, {Badenes}, {Kerzendorf},
  {Koester}, {Kromer}, {Miles}, {Pakmor}, {Schwab}, {Toloza}, {Toonen},
  {Townsley}, \& {Williams}}]{Shen2018}
{Shen}, K.~J., {Boubert}, D., {G{\"a}nsicke}, B.~T., {et~al.}
  2018{\natexlab{a}}, \apj, 865, 15

\bibitem[{{Shen} {et~al.}(2018{\natexlab{b}}){Shen}, {Kasen}, {Miles}, \&
  {Townsley}}]{Shen2018b}
{Shen}, K.~J., {Kasen}, D., {Miles}, B.~J., \& {Townsley}, D.~M.
  2018{\natexlab{b}}, \apj, 854, 52

\bibitem[{{Stasi{\'n}ska} {et~al.}(2004){Stasi{\'n}ska}, {Gr{\"a}fener},
  {Pe{\~n}a}, {Hamann}, {Koesterke}, \& {Szczerba}}]{2004A&A...413..329S}
{Stasi{\'n}ska}, G., {Gr{\"a}fener}, G., {Pe{\~n}a}, M., {et~al.} 2004, \aap,
  413, 329

\bibitem[{{Toal{\'a}} {et~al.}(2015){Toal{\'a}}, {Guerrero}, {Todt}, {Hamann},
  {Chu}, {Gruendl}, {Sch{\"o}nberner}, {Oskinova}, {Marquez-Lugo}, {Fang}, \&
  {Ramos-Larios}}]{2015ApJ...799...67T}
{Toal{\'a}}, J.~A., {Guerrero}, M.~A., {Todt}, H., {et~al.} 2015, \apj, 799, 67

\bibitem[{{Todt} {et~al.}(2015){Todt}, {Guerrero}, {Fang}, {Toala}, {Arthur},
  {Blair}, {Chu}, {Gruendl}, {Hamann}, {Marquez-Lugo}, {Oskinova}, {Ruiz},
  {Steffen}, \& {Schoenberner}}]{2015ASPC..493..141T}
{Todt}, H., {Guerrero}, M.~A., {Fang}, X., {et~al.} 2015, in Astronomical
  Society of the Pacific Conference Series, Vol. 493, 19th European Workshop on
  White Dwarfs, ed. P.~{Dufour}, P.~{Bergeron}, \& G.~{Fontaine}, 141

\bibitem[{{Tylenda} {et~al.}(1993){Tylenda}, {Acker}, \&
  {Stenholm}}]{1993A&AS..102..595T}
{Tylenda}, R., {Acker}, A., \& {Stenholm}, B. 1993, \aaps, 102, 595

\bibitem[{{Unglaub} \& {Bues}(2000)}]{2000A&A...359.1042U}
{Unglaub}, K. \& {Bues}, I. 2000, \aap, 359, 1042

\bibitem[{{van Regemorter}(1962)}]{1962ApJ...136..906V}
{van Regemorter}, H. 1962, \apj, 136, 906

\bibitem[{{Webbink}(1984)}]{Webbink1984}
{Webbink}, R.~F. 1984, \apj, 277, 355

\bibitem[{{Weidmann} {et~al.}(2023){Weidmann}, {Werner}, {Ahumada}, {Pignata},
  \& {Firpo}}]{2023A&A...676A...1W}
{Weidmann}, W.~A., {Werner}, K., {Ahumada}, J.~A., {Pignata}, R.~A., \&
  {Firpo}, V. 2023, \aap, 676, A1

\bibitem[{{Werner}(1992)}]{1992LNP...401..273W}
{Werner}, K. 1992, in The Atmospheres of Early-Type Stars, ed. U.~{Heber} \&
  C.~S. {Jeffery}, Vol. 401, 273

\bibitem[{{Werner} {et~al.}(2003){Werner}, {Deetjen}, {Dreizler}, {Nagel},
  {Rauch}, \& {Schuh}}]{2003ASPC..288...31W}
{Werner}, K., {Deetjen}, J.~L., {Dreizler}, S., {et~al.} 2003, in Astronomical
  Society of the Pacific Conference Series, Vol. 288, Stellar Atmosphere
  Modeling, ed. I.~{Hubeny}, D.~{Mihalas}, \& K.~{Werner}, 31

\bibitem[{{Werner} {et~al.}(1995){Werner}, {Dreizler}, {Heber}, {Rauch},
  {Wisotzki}, \& {Hagen}}]{Werner1995}
{Werner}, K., {Dreizler}, S., {Heber}, U., {et~al.} 1995, \aap, 293, L75

\bibitem[{{Werner} \& {Herwig}(2006)}]{2006PASP..118..183W}
{Werner}, K. \& {Herwig}, F. 2006, \pasp, 118, 183

\bibitem[{{Werner} \& {Rauch}(2014)}]{2014A&A...569A..99W}
{Werner}, K. \& {Rauch}, T. 2014, \aap, 569, A99

\bibitem[{{Werner} \& {Rauch}(2015)}]{2015A&A...584A..19W}
{Werner}, K. \& {Rauch}, T. 2015, \aap, 584, A19

\bibitem[{{Werner} {et~al.}(2004){Werner}, {Rauch}, {Reiff}, {Kruk}, \&
  {Napiwotzki}}]{2004A&A...427..685W}
{Werner}, K., {Rauch}, T., {Reiff}, E., {Kruk}, J.~W., \& {Napiwotzki}, R.
  2004, \aap, 427, 685

\bibitem[{{Werner} {et~al.}(2022){Werner}, {Reindl}, {Geier}, \&
  {Pritzkuleit}}]{2022MNRAS.511L..66W}
{Werner}, K., {Reindl}, N., {Geier}, S., \& {Pritzkuleit}, M. 2022, \mnras,
  511, L66

\bibitem[{{Zhang} {et~al.}(2019){Zhang}, {Fuller}, {Schwab}, \&
  {Foley}}]{2019ApJ...872...29Z}
{Zhang}, M., {Fuller}, J., {Schwab}, J., \& {Foley}, R.~J. 2019, \apj, 872, 29

\end{thebibliography}

\end{document}